\documentclass{aastex}

\newcommand\hb{H${\beta}$~}
\newcommand\ha{H${\alpha}$~}
\newcommand\nii{[N{\sc ii}]~}
\newcommand\sii{[S{\sc ii}]~}

\newcommand\oiii{[O{\sc iii}]~}

\newcommand\hii{H II~}

\slugcomment{}
\shortauthors{Stanghellini et al.}
\shorttitle{SMC \hii regions} 

\begin{document}

\title{{\it Hubble Space Telescope} observations of three
very young star clusters in the Small Magellanic Cloud.
\footnote{Based on observations made with the NASA/ESA Hubble Space Telescope, 
obtained at the Space Telescope Science Institute, which is operated 
by the Association of universities for research in Astronomy, Inc., under
NASA contract NAS 5--26555}}

\author{Letizia Stanghellini\altaffilmark{2}}
\affil{Space Telescope Science Institute, 3700 San Martin Drive,
Baltimore, Maryland 21218, USA; lstanghe@stsci.edu}

\author{Eva Villaver}
\affil{Space Telescope Science Institute; villaver@stsci.edu}

\author{Richard A. Shaw}
\affil{National Optical Astronomy Observatory, 950 N. Cherry Av.,
Tucson, AZ  85719, USA; shaw@noao.edu}

\and

\author{Max Mutchler}
\affil{Space Telescope Science Institute; mutchler@stsci.edu}

\altaffiltext{2}{Affiliated with the Space Telescope Division of the European Space 
Agency, ESTEC, Noordwijk, The Netherlands; 
on leave from INAF- Osservatorio Astronomico di Bologna}

\begin{abstract} 
We present {\it Space Telescope Imaging Spectrograph} (STIS) 
broad band imagery and optical slitless spectroscopy of three young
star clusters
in the Small Magellanic Cloud (SMC). MA~1796 and MG~2 were previously
known as Planetary Nebulae, and observed as such within our {\it Hubble
Space Telescope (HST)} 
survey.
With the {\it HST} spatial resolution, we show that they are instead H II regions,
surrounding very young star clusters. A third compact 
\hii~region, MA~1797, was serendipitously observed 
by us as it falls in the same frame of MA~1796. 
Limited nebular analysis is presented as derived from the
slitless spectra. We find that MA~1796 and MG~2 are
very heavily extincted, with c$\ge$1.4, defining them as the most extincted 
optically-discovered star forming regions in the SMC. 
MA~1796 and MG~2 are extremely compact (less than 1 
pc across), while MA~1797, with diameter of about 3 pc, is similar to the 
ultra compact \hii regions already known in the SMC.
Stellar analysis is presented, and approximate reddening correction for the stars is
derived from the Balmer decrement. Limited analysis of their stellar content and
their ionized radiation shows that these compact \hii
regions are ionized by small stellar
clusters whose hottest stars are at most of the B0 class.
These very compact, extremely reddened, and probably very dense 
\hii regions in the SMC offer insight in the most
recent star formation episodes in a very low metallicity galaxy.

\end{abstract}


\section {Introduction} 

The very compact \hii regions in the Magellanic Clouds, extremely reddened,
yet observable at the optical wavelengths, represent the early 
evolutionary stage of the cradles of the more
recent star formation (SF) \citep{2002ARA&A..40...27C}. 
If the duration of SF
episodes is related to the size of the SF regions \citep{2000ApJ...530..277E}, then
the very compact \hii regions in the SMC are the ideal probes of recent star formation
in a low metallicity environment.

While completing our {\it HST} snapshot survey of SMC Planetary Nebulae  (PNs;
{\it HST} Program 8663, Cycle 10; see Stanghellini et al. 2003), we found that two of the
target
PNs where 
misclassified in the ground-based catalog \citep{MA93}. On the basis of their
morphological properties in both the narrow and broad band images we found that
these targets are 
the most compact, most reddened \hii regions in the SMC known to date.
Furthermore, in one of the \hii region frames we serendipitously observed an
additional, very compact \hii region, previously classified
as an emission-line object 
(and possibly a \hii region) by \citet{MA93}.
Our broad band and slitless spectra acquired with STIS disclose some 
extreme properties of these three regions, including extreme extinction, that are worth discussing.

Only a handful of objects similar to the ones presented here
are currently known in the Magellanic Clouds \citep{2001A&A...372..667T},
and only three of them have been discovered in the SMC
\citep{2001A&A...372..667T,1985A&A...145..170T,1988A&A...195..230H}. The sample
presented in this paper then doubles the existing database of 
\hii regions that are much more compact
than the typical Magellanic Cloud \hii regions \citep{1990A&A...229..533C}, and that
have very high reddening. 

In this paper we present the {\it HST} stellar and nebular data of MA~1796, 
MA~1797, and MG~2. In $\S$2 we give an overview of the observing strategy.
In $\S$3 we discuss the imaging and spectral analysis, including the determination
of the extinction from the nebular Balmer emission, the ionizing flux inferred from 
nebular Balmer emission, the visual stellar magnitudes of the brightest stars
within the regions, and the cluster NIR magnitudes.
$\S$4 includes a discussion of our findings in the context of star formation 
in the SMC.

\section{Observations}

The observations were acquired with STIS. 
All observations were made with the CCD detector, in direct imaging
(50CCD)
and slitless mode. The spatial scale of the CCD is 0.051 arcsec pixel$^{-1}$,
corresponding to 0.014 pc pixel$^{-1}$
at the distance of the SMC. This 
allows a good spatial resolution to study the \hii region structures.
Each imaging observation was split in two, to
allow easy cosmic rays removal. The slitless spectra were acquired
with the G430M and G750M gratings. 
Observations with the G430M grating cover the range 4818 \AA~ to
5104 \AA~ at 0.28 \AA~ pixel$^{-1}$, and those with the G750M grating cover 
the range 6295 \AA~ to 6867 \AA~ at 0.56 \AA~ pixel$^{-1}$.
The exposures where originally planned to get good signal to noise ratio in the 
brightest emission lines of PNs, thus were not ideally designed with \hii regions in mind.

The observing log is reported in Table 1, where we list the targets, their
coordinates, the
observing date, the
data set name, the spectral element used in the observations, and the 
exposure time and number of exposures obtained. 

MA~1796 has been classified as 
a new SMC PN by \citet{MA93} (SIMBAD resolver: [MA93] 1796).
It is located on the outer side of the larger, young stellar cluster NGC~456. 
MA~1796 and NGC~456 could be physically associated, but our data are not helpful in 
establishing any connection.

MG~2 is catalogued as a PN
in \citet{1985MNRAS.213..491M} (SIMBAD resolver: MGPN SMC 2). There are no published
spectra of this object, nor images that resolve the \hii region spatially.

MA~1797 (SIMBAD resolver: [MA93] 1797) was previously 
catalogued as a possible \hii region by \citet{MA93}, possibly part of the DEM~152
cluster. 

The STIS data were calibrated using the standard
pipeline system, as in the LMC PN data \citep{sha01}.
Figure 1 show the observed \hii regions in the three
observing modes: broad band CCD, \oiii, and \ha images. The images are
to scale, to reflect the relative physical scales. The apparent and 
physical dimensions of the nebulae are listed in Table 2, columns (2) and (3). 

Spectral analysis have been performed in the same way to
that of LMC and SMC PNs \citep{sta02,sta03}. The combination of
dispersion and spatial scale allows a clear separation
of the monochromatic images for the major emission lines. 
We extract the
one-dimensional spectra and applied a photometric calibration 
using the standard STIS calibration pipeline module {\bf x1d} 
\citep{McGrath_etal99}.
We used extraction boxes for the regions large enough to encompass the
nebular features, but snug enough as to exclude most of the sky background
from the extraction. Sky background regions were selected for each object to avoid
stray stellar photons from field stars. The background was then averaged and
subtracted. 

We measured emission line intensities with IRAF\footnote{IRAF is distributed by
the National Optical Astronomical Observatory, which is operated by the 
Association of Universities for research in Astronomy, Inc., under cooperative
agreement with the National Science Foundation.} {\bf splot} task, 
fitting gaussians to individual lines, while estimating the continuum level. 
In the cases in which the emission lines were notably non gaussian, we 
estimated the line flux as measured from the area above the continuum.

The measured fluxes are listed in Table 2, columns (4), (6), and (7). 

We identified the \hb and the \oiii $\lambda\lambda$ 4959,5007
lines in the G430M spectrum of MG~2. 
The \ha emission is the only nebular feature in 
the G750M spectrum of this \hii region. 
The given \ha flux includes the \nii emission, although the 
slitless spectrogram suggest that the \nii contribution to the flux is marginal.

The analysis of the MA~1796 spectra shows the \oiii and \hb lines detected with 
a very good signal to noise ratio. The \sii emission at $\lambda$6716,6731 is also detected.

MA~1797 was serendipitously observed at
the edge of the field, so the \hb emission does not appear in the G430M
spectrum. 
The \ha flux is high,
comparable with that of the previously observed compact \hii regions N~81 and N~88A.

\section{Data analysis and physical properties}

\subsection{Nebular reddening}

We derive the optical extinction from the analysis of the nebular Balmer lines, and we
give the logarithmic extinction constants (c) in Table 2, column (5). We assume that the optical
extinction law is the same in the Galaxy and the SMC, thus we do not estimate the foreground
extinction separately. The derived extinction is rather a lower limit, since
we do not consider possible regions whose optical flux is totally absorbed. 
Our feeling is that the extinction corrections should be higher than what estimated
from the nebular lines.

With a \ha/\hb flux ratio of 8.7, MG~2 is evidently an extremely reddened object. 
Note that the flux ratio could be off by a factor of two, given the flux uncertainties, 
and still be extremely high for a compact \hii regions.  
By assuming T$_{\rm e}$=10,000 K, we estimate the logarithmic optical extinction 
constant c=1.4 (or E$_{\rm B-V}$=0.95, Seaton 1979). 
 
The optical extinction constant of MA~1796, c=1.53, is the highest ever measured in a 
compact SMC H~II region. This nebula is a factor of 10 lower in excitation than MG~2.

Given that we do not measured its \hb flux, we could not derive the extinction
constant for MA~1797.

\subsection{Ionizing flux from the nebular Balmer emission}

We computed the RMS density of MA~1796 and MG~2, based upon our knowledge 
of the \hb flux, the average nebular extinction, the known distance 
to the SMC, and the physical extent of the nebulae, using the formulation:

$${\rm N}_{\rm RMS}= [({\rm F}_{\beta} / {\rm h}\nu~\alpha) ({\rm W/V})]^{0.5}\eqno(1),$$

where $\nu$ is the 
\hb (or \ha) frequency, $\alpha$ is the recombination coefficient for \hb (or \ha),
V is the nebular volume, and W is the geometrical dilution factor.  
The log density is 
given in column (8) of Table 2.  We also estimated the total number of 
ionizing photons, Q$_{\rm H}$, for the Str\"omgren spheres corresponding to the sizes 
of these nebulae \citep{ost89}, assuming that all ionizing radiation is
contained within the regions.  
The results, in the last column of Table 3, are roughly 
consistent with a single B0 or a very late O main-sequence star exciting the region
\citep{ost89}.  The value of Q$_{\rm H}$ is very uncertain because 
it is proportional to the square of the electron density, which itself 
is very uncertain.  It also depends upon the assumption of high optical 
depth to ionizing radiation, which is probably a reasonable approximation 
for ultra compact \hii regions. We did not attempt an estimate of the
fraction of the ionizing radiation absorbed by dust (see discussion of the reddening).

In the case of MA~1797 we perform the calculation
of the density and Q$_{\rm H}$ from the \ha flux, uncorrected for extinction.
The resulting analysis for this region is extremely uncertain, and it is given in Table 2 
only as an indication. In the next section we make use of these 
approximate values for Q$_{\rm H}$ in these nebulae to aid our interpretation of 
the stellar photometry.

\subsection{Visual stellar photometry}

With the high spatial resolution of {\it HST}
we can easily resolve the stars individually, thus we can use aperture 
photometry with confidence. 
We have measured the magnitudes of the brightest stars within MA~1796 and MG~2.
The stellar photometry was performed with
the {\it apphot} package in IRAF, through a circular aperture of 2 pixels
of radius. The averaged sky within an annulus 1 pixel wide 
adjacent to the position of the star
was then subtracted to the stellar measurement. The sky includes the nebular 
emission as well. The measured instrumental
magnitudes (STMAG system) use the zero-point calibration by \cite{Betal:02}. 
The aperture
correction has been taken from the curve of encircled energy derived by
\cite{Betal:02} for stars near the field center. 

The 50CCD mirror has broad-band transmission, and its response curve is far 
from that of the standard V filter. 
The transformation between the measured instrumental magnitudes and the
standard V magnitudes depends on the spectral energy distribution (SED) of the
source, which is determined by the 
amount of extinction and, to a lesser extent, by the stellar temperature. 
We do not have the
stellar temperatures of the stars, but we know that they are
hot enough to ionize \hii regions, thus we infer their temperatures to be
in the range between 30,000 and 50,000 {\rm K}. We 
assume that the extinction toward the stars is the average nebular
extinction. We feed these parameters to the 
IRAF/STSDAS routine {\bf synphot}, to perform the synthetic photometric modeling 
necessary to derive the transformation to the standard V
magnitudes.

In Table 3 we give the photometric measurements 
of the three brightest stars in MA~1796, and the 5 
brightest stars in MG~2. The quoted errors 
in the V standard magnitudes include the standard deviation of the transformation 
for the range of temperatures used.
In Figures 2 and 3 we 
identify the stars with the same numbering than 
in Table 3, for MA~1796 (Fig. 2) and MG~2 (Fig. 3).
We have corrected the V magnitudes for extinction
by using the
approximate relation between the average logarithmic extinction
constant derived for the nebula, c, and the color excess,
E$_{B-V}$ ({\rm c}~=~1.47~E$_{B-V}$, Seaton 1979). The
SMC extinction law is very similar to the Galactic law in this
wavelength range \citep{Hut:82}, so we have used the
Galactic reddening curve law of \cite{Sm:79}, and assumed
that R$_{V}$~=~3.1. The stellar extinction in magnitudes we
apply is then A$_{V}$~=~2.1~c.

In order to
derive the absolute visual magnitudes, we have used the 
average SMC distance modulus obtained with the RR Lyrae 
method \citep{wes97}. 
None of the stellar magnitudes measured and
transformed as described above correspond to an O star
of any subclass, according to the calibration of \cite{Vgs:96}, the 
explanation probably lies in the extinction, being higher than measured.

\subsection{Cluster NIR magnitudes}

In order to frame the extinction correction a little better,	     
we have searched the two Micron All Sky Survey (2MASS) for near Infrared
sources within 5" of the central position of 
MA~1796 and MG~2. We found that each H II region is uniquely associated with 
one compact object in the 2MASS point source
catalogue (All-Sky Point Source Catalog). The J, H, and Ks
magnitudes corresponding to these two point sources are listed In Table 4.

In a JHK color-color diagram (e.g. \citealt{La:92}) MA~1796 and
MG~2 would lay outside and to the right of the reddening band, indicating the
presence of intrinsic infrared excess. This result alone could have 
indicated that the objects are not common PNs. Intrinsic infrared excesses
are indicative of the presence of circumstellar material. According to
\citet{La:92}, young stellar
objects and Herbig Ae/Be stars populate the part of the Near-Infrared
color-color diagram where our sources are located.

\section{Summary}

We have identified two new compact young stellar cluster in the SMC, that were
previously classified as PNs.
The unique spatial resolution 
of {\it HST} clearly disclose that these objects, merely a few arcsecond across,
are not PNs. 

Limited analysis of their nebular spectra
disclose very high \ha to \hb ratios,
placing them among the extreme compact and reddened H II regions
in the SMC, as observed at the optical wavelengths.
Imaging and the \ha flux are also available for a third, 
larger H II region that was serendipitously observed in the same frame of
one of the misclassified PNs. 

Studying the IR properties of the regions,
we found that they present high IR excess. These excess, 
together with compactness of the regions, indicate that they are
very young star forming regions.

Stellar photometry places the brightest stars in the 
regions out of the O star domain. The definitive stellar classification is
not possible with the present data. 
If the stars observed near the center of these nebulae are in fact 
providing all or most of the ionizing photons, then the spectral 
types inferred from the observed brightness and the bolometric
correction are inconsistent 
with the observed level of nebular ionization.  A possible explanation 
is that the stars suffer greater extinction than is inferred from the 
nebular Balmer decrement. 

We conclude that we observed a few very young clusters of recent star formation,
but the details of the populations therein are left to a future investigation.
We plan to perform UBV photometry of these regions, possibly with {\it HST} 
resolution, to disclose the true nature of the component stars.
Infrared observations at high resolution are also necessary to
analyze in detail the reddening structure of these unique H II regions.
A small cluster of stars whose few brightest and hottest members are of 
early B or late O spectral type is consistent both with the level of nebular ionization, 
and with the small physical size of the nebula. 

We also plan to obtain the physical parameters of the gas, such as 
temperature, density, and abundances, from moderate resolution ground based
optical spectra. Such observations will also verify our estimated 
of the ionizing photons from the embedded stars.

This work was supported by NASA through grant GO-08663.01-A from Space
Telescope Science Institute, which is operated by the Association of 
Universities for Research in Astronomy, Incorporated, under NASA contract
NAS -26555. 
This publication made use of data products from the Two Micron All Sky
Survey, which is a joint project of 
the University of Massachusetts and the Infrared Processing and Analysis
Center/California Institute of 
Technology, funded by the National Aeronautics and Space Administration and
the National Science Foundation.

\clearpage
\begin{deluxetable}{lrrclcrc}

\tablewidth{16truecm}
\tablecaption {Observing log \label{ObsLog}}
\tablehead {
\colhead {} & \colhead {R.A.} & \colhead {Dec.} & \colhead {} & 
\colhead {} & \colhead {} & \colhead {t$_{Exp}$} & \colhead {} \\
\colhead {Name} & \colhead {(J2000)\tablenotemark{a}} & \colhead {(J2000)\tablenotemark{a}} & \colhead {Date} & 
\colhead {Data set} & \colhead {Disperser} & \colhead {(s)} & \colhead {N$_{Exp}$} 
}
\startdata  
MA 1796 &  1:14:47.26 &  $-73$:20:14.2 & 2000-Sep-27 & O65S26010 & MIRVIS & 120 & 2 \\*
       & & &            & O65S26020 & G750M & 100 & 2 \\*
       & & &               & O65S26030 & G750M & 400 & 2 \\*
       & & &               & O65S26040 & G430M & 200 & 2 \\
       & & &               & O65S26KGQ & MIRVIS & 15 & 1 \\
MA 1797 \tablenotemark{b} &  1:14:47.17 &  $-73$:19:46.7 & 2000-Sep-27 & O65S26010 & MIRVIS & 120 & 2 \\*
       & & &               & O65S26020 & G750M & 100 & 2 \\*
       & & &               & O65S26030 & G750M & 400 & 2 \\*
       & & &               & O65S26040 & G430M & 200 & 2 \\
       & & &               & O65S26KGQ & MIRVIS & 15 & 1 \\
MG 2 &  0:28:10.13 &  $-72$:58:44.1   & 2000-Nov-7   & O65S28010 & MIRVIS & 300 & 2 \\*
       & & &               & O65S28020 & G750M & 1300 & 2 \\*
       & & &               & O65S28030 & G430M & 800 & 2 \\
\enddata  
\tablenotetext{a}{Coordinates measured from data.}
\tablenotetext{b}{MA 1797 was serendipitously observed in the MA 1796 field (same exposures).}
\end{deluxetable}

\clearpage

\begin{deluxetable}{lllllllll}
\tablewidth{20truecm}
\tablecaption {Nebular analysis\label{Flux}}
\tablehead {
\colhead {Name} & \colhead {Size} & \colhead{Radius} & \colhead {log F$_{\rm (H\beta)}$} & \colhead {c} & 
\colhead {log F$_{\rm [O III]\lambda5007}$} & \colhead {log F$_{\rm H\alpha}$\tablenotemark{a}} & \colhead{log N$_{\rm RMS}$} & \colhead{log Q$_{\rm H}$}  \\
\colhead {} & \colhead {[\arcsec]} & \colhead{pc}& \colhead {[erg cm$^{-2}$ s$^{-1}$]} & \colhead {} & 
\colhead {[erg cm$^{-2}$ s$^{-1}$]}  & \colhead {[erg cm$^{-2}$ s$^{-1}$]} & \colhead [cm$^{-3}$] & \colhead{}\\
}
\startdata  
MA 1796  & 3.0  &  0.424& -13.85$\pm$0.25 & 1.53& -13.71$\pm$0.25&  -12.85$\pm$0.2 & 2.61& 47.61\\
MA 1797  & 11.0 &  1.554&  \nodata  & \nodata & \nodata  & -11.38$\pm$0.15 & 2.07& 48.23 \\
MG 2     & 3.5  &   0.495& -14.35$\pm$0.25& 1.4 & -13.14$\pm$0.2 & -13.41$\pm$0.2 & 2.20& 46.98\\
\enddata  
\tablenotetext{a}{Intensity of the blend of \ha and \nii 
$\lambda$6548, 6584.}

\end{deluxetable}

\clearpage

\begin{deluxetable}{lccc}
\tablewidth{8truecm}
\tablecaption {Stellar photometry}
\tablehead {
\colhead {Star} & \colhead {ST\tablenotemark{a}} & \colhead {V} & \colhead {M$_{\rm V}$} \\

\colhead {} & \colhead {mag} & \colhead {mag} & \colhead {mag}\\ 
}
\startdata  
MA~1796&&&\\

\#1&       21.62$\pm$0.11  &     18.28   &  -0.55 \\
\#2&       21.73$\pm$0.11  &     18.38   &  -0.45 \\
\#3&       21.11$\pm$0.05  &     17.76   &  -1.07 \\
Cluster\tablenotemark{b}&       18.14$\pm$0.01  &     14.79   &  -4.04 \\
			
MG~2&&&\\
            
\#1 &   22.31$\pm$0.06  &   19.21  &  0.38   \\     
\#2&   23.23$\pm$0.40  &   20.13  &  1.3	  \\
\#3 &   22.56$\pm$0.27  &   19.46  &  0.63	  \\
\#4&   22.48$\pm$0.03  &   19.38  &  0.55	  \\
\#5&   22.72$\pm$0.10  &   19.62  &  0.79	  \\
Cluster\tablenotemark{c}&   19.55$\pm$0.013 &   16.45  & -2.37	  \\

\enddata  

\tablenotetext{a}{Uncorrected for extinction}
\tablenotetext{b}{Including the central 1.5 \arcsec}
\tablenotetext{c}{Including the central 1.0 \arcsec}
\end{deluxetable}

\clearpage

\begin{deluxetable}{lccc}
\tablewidth{10truecm}
\tablecaption {Cluster infrared photometry}
\tablehead {
\colhead {Cluster} & \colhead {J} & \colhead {H} & \colhead {K$_{\rm S}$} \\

\colhead {} & \colhead {mag} & \colhead {mag} & \colhead {mag}\\ 
}
\startdata  

MA~1796&  15.53$\pm$0.08& 14.96$\pm$0.10 & 14.24$\pm$0.08 \\
MG~2&     16.11$\pm$0.11& 15.63$\pm$0.16 & 14.54$\pm$0.10 \\
			
\enddata  
\end{deluxetable}

\clearpage

\figcaption{broad- and narrow-band images of the new SMC \hii regions. From
top left, clockwise: MA~1796 broad band image, MA~1796 and MG~2 broad,
\oiii $\lambda$ 5007, and \ha images, and MA~1797 \ha, 
\oiii $\lambda$ 4959 and 5007 images.}

\figcaption{Central part of the MA~1796 cluster, with stars
labeled as in Table 3}
\figcaption{Central part of the MG~2 cluster, with stars
labeled as in Table 3}

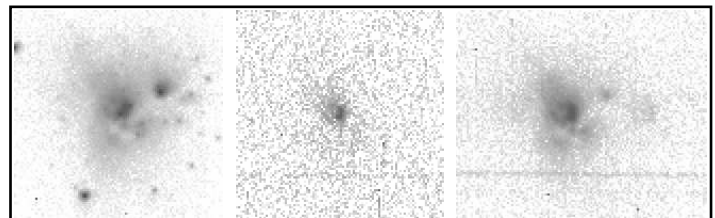
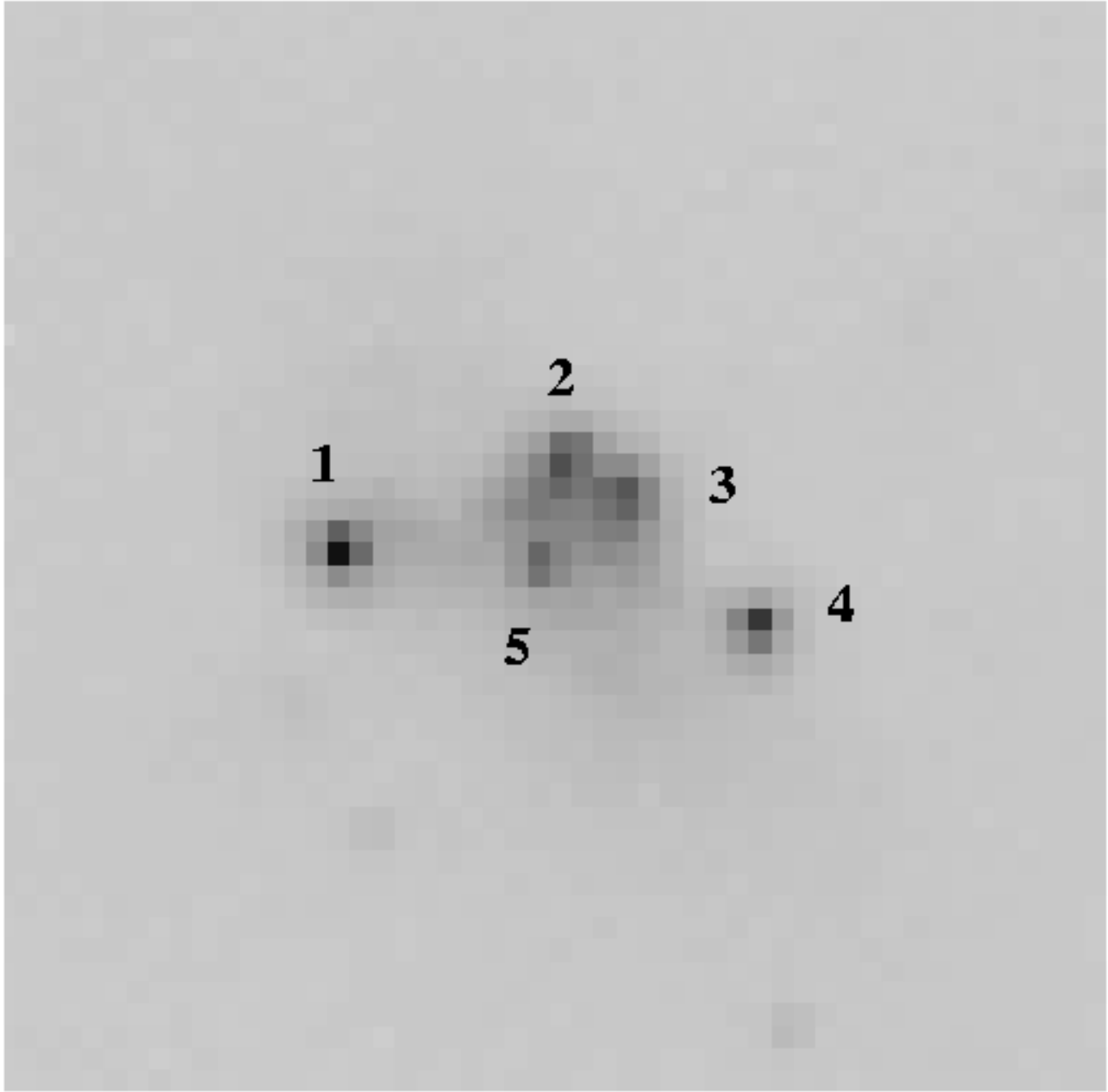
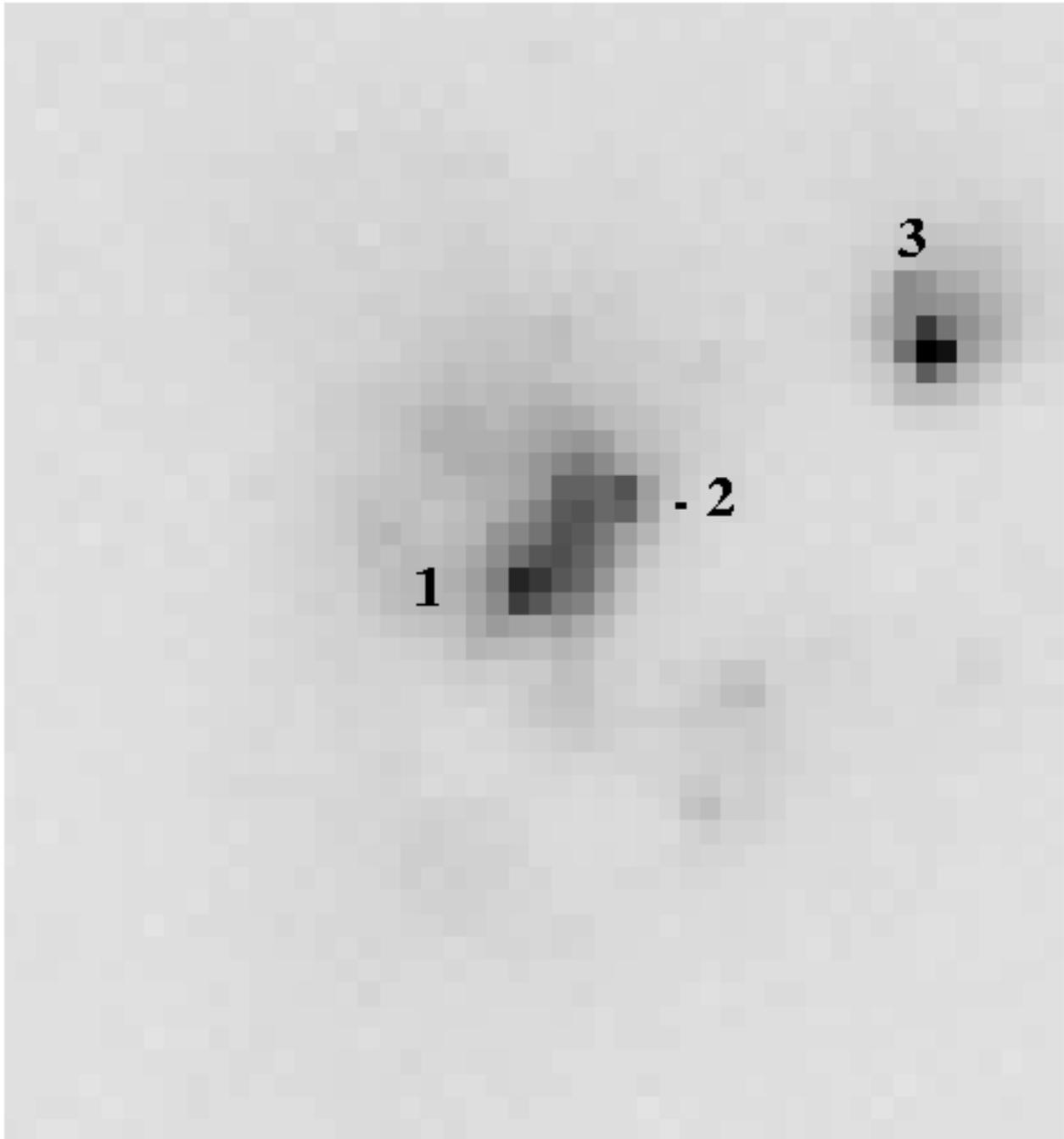
\end{document}